# Giant Goos-Hänchen shift in Scattering: the role of interfering Localized Plasmon modes

J. Soni,[1] S. Mansha,[2] S. Dutta Gupta,[3,*] A. Banerjee,[1,*] and N. Ghosh[1,*]

[1]*Dept. of Physical Sciences, Indian Institute of Science Education and Research (IISER) Kolkata, Mohanpur Campus, Mohanpur 741 252, India*
[2]*Division of Physics and Applied Physics, School of Physical and Mathematical Sciences, Nanyang Technological University, Singapore, 637371 Singapore*
[3]*School of Physics, University of Hyderabad, Hyderabad 500046, India*
*Corresponding authors: sdghyderabad@gmail.com, ayan@iiserkol.ac.in, nghosh@iiserkol.ac.in*

The longitudinal and the transverse beam shifts, namely, the Goos-Hänchen (GH) and the Spin-Hall (SH) shifts are usually observed at planar interfaces. It has recently been shown that the transverse SH shift may also arise due to scattering of plane waves. Here, we show that analogous in-plane (longitudinal) shift also exist in scattering of plane waves from micro/nano systems. We study both the GH and the SH shifts in plasmonic metal nanoparticles/ nanostructures and dielectric micro-particles employing a unified framework that utilizes the transverse components of the Poynting vector of the scattered wave. The results demonstrate that interference of neighboring resonance modes in plasmonic nanostructures (e.g., electric dipolar and quadrupolar modes in metal spheres) leads to giant enhancement of GH shift in scattering from such systems. We also unravel interesting correlations between these shifts with the polarimetry parameters, *diattenuation* and *retardance*.

The longitudinal and the transverse shifts of a physical light beam at planar dielectric interfaces, namely, the Goos-Hänchen (GH) and the Imbert-Federov (IF) shifts have been investigated for a long time [1-5]. These in-plane (in the plane of incidence) and transverse (perpendicular to the plane of incidence) shifts have been typically observed in reflection/refraction, total internal reflection of light beam at planar interfaces and have accordingly been modeled using various theoretical treatments [6-13]. While the GH shift originates from the angular gradient of the complex reflection / refraction coefficients, the IF shift has its origin in spin orbit interaction (SOI) and conservation of total angular momentum [4,5, 11-13]. The latter also leads to the so-called Spin Hall (SH) effect of light [4,8, 11-13]. These shifts depend on the polarization state of the incident beam; the eigenmodes of the GH shift are TM (p) and TE (s) linearly polarized waves and that of the SH shift are left and right circularly polarized (LCP and RCP) waves respectively [4,5]. Note that the IF shift is more general than the SH shift in the sense that its eigenmodes are both left/right circular polarization states and +45º/-45º linear polarization states [4]. It has recently been shown that the transversal spin transport (transverse SH shift) may also arise due to scattering of plane waves from objects having higher dimensionality (e.g., sphere) [14]. The question therefore naturally follows – is there an analogous in-plane (longitudinal) shift in scattering of plane waves? We show here that indeed such longitudinal GH shift can be mediated in scattering of plane waves from micro/nano systems. We study both the scattering-mediated GH and SH shifts utilizing the transverse components of the Poynting vector of the scattered wave. The role of the plane of incidence (for planar interface) is played here by the scattering plane. Consequently, the two transverse components of the Poynting vector of the scattered wave, the in-plane (along the direction of the polar scattering angle) and the out of plane (along the direction of the azimuthal angle) components are responsible for the longitudinal (in the plane of scattering) and the transverse (perpendicular to the plane of scattering) shifts respectively. Importantly, we demonstrate that interference of neighboring localized plasmon resonance modes in metal nanostructures / nanoparticles leads to giant enhancement of GH shift. We also derive interesting correlations between the GH and the SH shifts with scattering polarimetry parameters, namely, *retardance* $\delta$ (phase shift between orthogonal polarization states) and *diattenuation* $d$ (differential attenuation of orthogonal polarization states) [15,16].

In order to derive expressions for the GH and the SH shifts, we consider scattering of plane waves by a spherical particle [17]. We chose Cartesian coordinate system with the incident plane wave propagating in the $Z$ direction, the two orthogonal axes $X$ and $Y$ representing the polarization axes. The polar angle of scattering $\theta$ and the azimuthal angle $\phi$ are assigned with respect to the Z and the X-axes respectively. Since, the eigenmodes of the GH shift are linearly polarized waves, we first consider scattering of X-polarized wave. The radial, angular and the azimuthal components of the scattered electric and magnetic fields can be derived using Mie theory [17]. These can then be utilized to obtain the corresponding radial ($P_r$), angular ($P_\theta$) and azimuthal ($P_\phi$) Poynting vector components. Note that the radial component of the Poynting vector signifies energy flow along the direction of propagation of the scattered wave, whereas, both the angular and the azimuthal components imply transversal energy flow and accordingly lead to shifts (parallel and perpendicular to the scattering plane respectively) in perceived location of the source (scatterer) [14,18,19]. These may be interpreted as the longitudinal GH and the transverse SH shifts mediated by the scattering process. Apparently, non-zero magnitudes of the radial components of the electric and the magnetic fields in the near field (or in the intermediate zone, depending upon its strength) [14], are the origin of these shifts. The GH shift in scattering (for X- polarized incident wave) can thus be defined using the angular component of the Poynting vector as

$$\vec{\Delta}_{GH} = \lim_{r\to\infty} r\left(\frac{P_\theta \hat{\theta}}{|P_r|}\right) \qquad (1)$$

$$\Delta_{GH} = -\frac{1}{k}\left(\frac{\mathrm{Im}\left(S^{2*}\left[\sum_{n=1}^{\infty}(2n+1)a_n\pi_n\right]\cos^2\varphi - S_1\left[\sum_{n=1}^{\infty}(2n+1)b_n\pi_n\right]^*\sin^2\varphi\right)}{|S_1|^2\sin^2\varphi + |S_2|^2\cos^2\varphi}\right)\sin\theta \qquad (2)$$

Here, $\rho = kr$, $k$ is the wave vector in the medium and $r$ is the radial distance; $a_n$ and $b_n$ are the well known coefficients of the normal scattering modes (TM (electric) and TE (magnetic) modes respectively) and $\tau_n$, $\pi_n$ are the corresponding angle ($\theta$) dependent functions; $S_2$ and $S_1$ are the amplitude scattering matrix elements [17], defined as

$$S_1 = \sum_{n=1}^{\infty}\frac{(2n+1)}{n(n+1)}(a_n\pi_n + b_n\tau_n), S_2 = \sum_{n=1}^{\infty}\frac{(2n+1)}{n(n+1)}(a_n\tau_n + b_n\pi_n) \qquad (3)$$

Note that the expressions for the GH shifts $\Delta_{GH}$ for X, Y linear polarization states (henceforth noted as $|H\rangle$ and $|V\rangle$ states) would exhibit symmetry with respect to the azimuthal angle (shift for $|H\rangle$ state at $\varphi=0$ is equal to the shift for $|V\rangle$ state at $\varphi=\pi/2$ and vice versa). In general, the GH shift depends upon both the incident linear polarization state and on azimuthal angle $\phi$ (orientation of the scattering plane). Several interesting trends can be predicted from the expression for shift. (a) For scattering dominated by one particular type of modes (either TM modes with $b_n \sim 0$; or TE modes with $a_n \sim 0$), shift corresponding to one of the linear polarization states (e.g., $|H\rangle$ state, $a_n$ mode) at a chosen scattering plane ($\phi = 0$) would be significant whereas the other ($|V\rangle$ state, $b_n$ mode at $\phi = 0$) would be negligible. We thus treat these as shifts introduced by the TM and TE scattering modes respectively. (b) In such a case, the magnitude of the shift would be determined by the interference of the neighboring modes. For example, for small metal spheres (radius $a<<\lambda$), scattering is primarily dominated by the lowest order TM modes (e.g., electric dipolar $a_1$ and quadrupolar $a_2$ modes) [15,20]. The magnitude of the GH shift may be enhanced by the interference of the neighboring plasmon resonance modes in metal nanoparticles / nanostructures. For small dielectric Rayleigh scatterers ($a<<\lambda$) on the other hand, where scattering can be treated in dipole approximation (only the first TM $a_1$ mode contributes), the angular component of the Poynting vector ($P_\theta$) would be $\sim$ zero, and accordingly the GH shift would vanish.

As previously noted, while the eigenmodes of IF shift are both LCP/RCP states and +45°/-45° linear polarization states, the eigenmodes of SH shift are solely the LCP/RCP states [4]. In order to derive expressions for the SH shift, we thus consider scattering of LCP / RCP incident wave. The radial, angular and azimuthal components of the Poynting vector for LCP / RCP wave can also be obtained employing Mie theory. This yields same magnitude and opposite signs of azimuthal component of the Poynting vector ($P_\phi$) for the LCP/RCP states. In contrast, the angular component ($P_\theta$) is independent of the polarization state (LCP / RCP). We thus define the SH shift as the circular polarization-dependent transverse (to the scattering plane and along the azimuthal direction) shift in the perceived location of the source, via the azimuthal component of the Poynting vector as

$$\vec{\Delta}_{SH} = \lim_{r\to\infty} r\left(\frac{P_\phi \hat{\phi}}{|P_r|}\right) \qquad (4)$$

$$\Delta_{SH} = \frac{\sigma}{k}\left(\frac{\mathrm{Re}\left(S_1^*\left[\sum_{n=1}^{\infty}(2n+1)a_n\pi_n\right] + S_2\left[\sum_{n=1}^{\infty}(2n+1)b_n\pi_n\right]^*\right)}{|S_1|^2 + |S_2|^2}\right)\sin\theta \qquad (5)$$

Where, $\sigma = \pm 1$ for LCP and RCP states respectively.

The differences in the expressions and the nature of the two shifts are worth a brief mention here. Unlike the GH shift, for the LCP/RCP states, the SH shift has the same magnitude and opposite signs and are independent of $\phi$ (choice of scattering plane). The shift is determined by both the strength of the individual scattering modes and the interference of different modes (e.g., $|a_1|^2, |a_2|^2$ and $\mathrm{Re}(a_2^* a_1)$ when two TM modes contribute). Thus, the effect of the interference of the modes is expected to be weaker on SH shift as compared to the GH shift. Moreover, unlike the GH shift, the SH shift would exist even when the contribution is from a single mode (e.g., $a_1$ mode) [18,19]. Although the above formalism is for spherical scatterer, extension for non-spherical particles is warranted. For example, for studying the shifts for rods, spheroids etc., the T-matrix approach [16] (instead of Mie theory) may be used to compute the scattering matrix elements. In what follows, we study the role of the interfering modes on both the GH and the SH shifts with illustrative example of resonant localized plasmon modes (TM- dipolar $a_1$ and quadrupolar $a_2$ modes) in metal nanospheres ($a<<\lambda$), and non-resonant higher order TM and TE ($a_n$ and $b_n$) modes in dielectric microspheres ($a\geq\lambda$).

**Figure 1a** shows the GH shifts ($\Delta_{GH}$, Eq. 2) in scattering from a metal (silver, Ag) nanosphere of radius $a$ = 50 nm with surrounding medium as water. The scattering angle ($\theta$) dependence of $\Delta_{GH}$ in the scattering plane $\phi = 0$ for incident X-polarized ($|H\rangle$ state) and Y-polarized ($|V\rangle$ state) light are shown in the figure and its inset respectively. The results are shown for two wavelengths, $\lambda$ = 380 nm and 430 nm. The observed trends can be summarized as – (a) The magnitudes of $\Delta_{GH}$ for $|V\rangle$ polarization state (for $\phi = 0$) are much weaker than that observed for $|H\rangle$ state. (b) The shifts exhibit strong dependence on scattering angle and wavelength. For incident $|H\rangle$ state, $|\Delta_{GH}|$ achieves its maximum value at $\theta \sim 62°$ for $\lambda = 380$ nm, whereas the corresponding maximum for 430 nm wavelength is at $\theta \sim 90°$. (c) Interestingly, the $|H\rangle$ polarized light exhibits a giant shift at 380 nm for $\theta \sim 62°$ ($|\Delta_{GH}| \sim 6\lambda$) (usually the GH shift is in the sub-wavelength domain). Moreover, the signs of $\Delta_{GH}$ are reversed [21] between 380 nm and 430 nm wavelength.

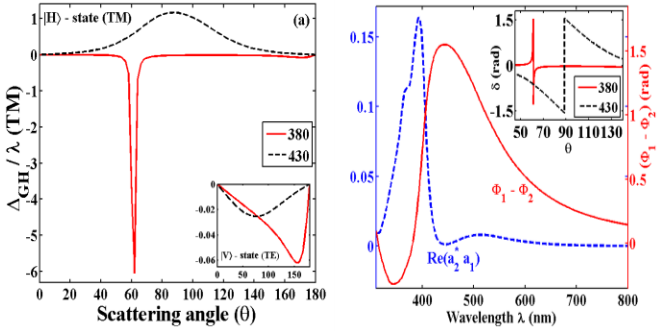

**Fig.1.** **(a)** Scattering angle θ dependence of the computed GH shift $\Delta_{GH}$ for Ag nanosphere (a=50 nm), at two wavelengths, λ=380 nm (red solid line), 430 nm (black dotted line). The shifts are shown for incident X-polarized light ($|H\rangle$ state) in the scattering plane $\phi = 0$. The inset shows the corresponding shifts for incident Y-polarized light ($|V\rangle$ state). **(b)** The wavelength dependence of the strength of the interference of the dipolar $a_1$ and the quadrupolar $a_2$ plasmon modes ($\mathrm{Re}(a_2^* a_1)$) (blue dotted line, left axis) and the corresponding phase difference (in radian) between them (red solid line, right axis). The inset shows the variation of retardance parameter of scattering $\delta$ (in radian) as a function of θ at two wavelengths λ = 380 nm and 430 nm for the same Ag nanosphere.

Note that the 50 nm Ag sphere exhibits two prominent plasmon resonances corresponding to the electric dipolar ($a_1$) and the quadrupolar ($a_2$) modes with resonance peaks at ~ 500 nm and 380 nm respectively [14,19]. The negligible contributions of the magnetic or TE modes ($b_n$ ~ 0) leads to weak magnitudes of the GH shift for $|V\rangle$ polarized light (Fig.1a, inset). As previously anticipated, the observed giant enhancement of the shift at λ = 380 nm (spectral overlap region of the two resonances) for incident $|H\rangle$ polarization state is a manifestation of the interference of the two neighboring plasmon resonance modes. This can be comprehended from **Figure 1b**, where the wavelength dependence of the strength of the interference of the two plasmon modes ($\mathrm{Re}(a_2^* a_1)$), and the corresponding phase difference between them are shown. The interference appears to be maximal at λ ~ 380 nm and decays sharply beyond 400 nm (significantly weak at 430 nm). Accordingly, the magnitude of the shift is significantly larger at 380 nm as compared to 430 nm (Fig. 1a). The apparent reversal of phase differences explains the observed reversal of signs of $\Delta_{GH}$ between 380 nm and 430 nm. Since the enhancement of the GH shift is controlled by the interference of the modes and the phase difference between them, it might be useful to derive correlations between the shift and the experimentally measurable scattering polarimetry parameter, *retardance* $\delta$. The parameter $\delta$ also deals with relative phases of the scattered wave and can be defined as [15,16]

$$\delta(\theta) = \tan^{-1}\left[\frac{\mathrm{Im}(S_2^*(\theta)S_1(\theta))}{\mathrm{Re}(S_2^*(\theta)S_1(\theta))}\right] \quad (6)$$

The scattering angle dependence of $\delta$ at two wavelengths λ = 380 nm and 430 nm for this *Ag* nanosphere is shown in the inset of Figure 1b. Comparison of $\delta(\theta)$ and $\Delta_{GH}(\theta)$ (Fig. 1a) reveals that large magnitude of GH shift is associated with sharp variation of $\delta$ with θ. For example, a rapid change in $\delta$ when θ approaches 62° for λ = 380 nm is accompanied with large value of $|\Delta_{GH}|$ ~ 6λ, whereas relatively smoother variation of $\delta$ when θ approaches 90° for λ = 430 nm leads to smaller value for $|\Delta_{GH}|$.

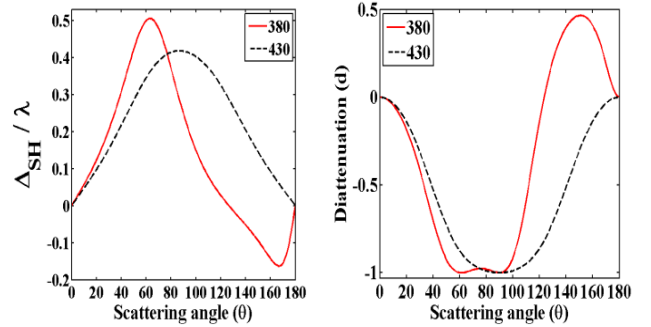

**Fig. 2. (a)** The computed SH shift $\Delta_{SH}$ as a function of θ for the same 50 nm Ag nanosphere of Fig. 1, for two wavelengths, λ=380 nm (red solid line), 430 nm (black dotted line). **(b)** The angular variation of *diattenuation d* for the Ag nanosphere, at λ=380 nm (red solid line) and 430 nm (black dotted line).

In **Figure 2a,** we show the computed (using Eq. 5) SH shift $\Delta_{SH}$ as a function of θ for the same Ag nanosphere as that of Fig. 1. As predicted, excitation of one particular type of modes (TM plasmon modes $a_1$ and $a_2$) does not lead to significant enhancement of the shift ($|\Delta_{SH}| < \lambda$). Moreover, the influence of the interference of these modes on the resulting SH shift is also weaker: $|\Delta_{SH}|$ at θ ~ 62° for λ = 380 nm (strong interference region) is only marginally larger than that observed at θ ~ 90° for λ = 430 nm (weak interference region). Interestingly, the angular variation $\Delta_{SH}(\theta)$ exhibits distinct correlation with the other scattering polarimetry parameter, namely, *diattenuation d*(θ) (shown in **Figure 2b**). The parameter $d$ can be defined as [15,16]

$$d(\theta) = \left\{\frac{|S_2(\theta)|^2 \cos^2\theta - |S_1(\theta)|^2}{|S_2(\theta)|^2 \cos^2\theta + |S_1(\theta)|^2}\right\}, \quad (7)$$

The correlation between $\Delta_{SH}$ and $d$ arises because both these parameters are related to spin orbit interaction of light and the resulting transformation of spin angular momentum (SAM) to orbital angular momentum (OAM) [14,15,22]. Note that for incident RCP/LCP state, the strength of SAM to OAM conversion depends upon the magnitude of $d$ ($d = 1$, corresponds to complete conversion, the scattered light becomes completely linearly polarized subsequently generating OAM) [15]. Since, the SH shift is also known to be large when the transformation of SAM to OAM is complete [14], local angular maxima of $|\Delta_{SH}|$ is accompanied with local maxima of $|d|$.

Till now, we presented results pertaining to the plasmonic metal nanoparticles. We have carried out similar studies for dielectric scatterers as well. **Figure 3a** shows the computed GH shift as a function of θ for a dielectric microsphere ($a = 1$ μm, refractive index $n_s = 1.6$, λ = 632.8 nm) with surrounding medium as air. For this large sized dielectric scatterer (a ≥ λ), several higher order TM and TE ($a_n$ and $b_n$) modes contribute in scattering, none of which may be resonant. $\Delta_{GH}$ exhibits rather complex $\theta$ dependence with its magnitude peaking at several narrow range of angles $\theta$. This apparently arises due to the complicated nature of the angular dependences of the contributing higher order modes and the resulting interferences. Despite such complexities, $\Delta_{GH}(\theta)$ exhibits similar correlation with the retardance parameter $\delta$ (like that of the plasmonic nanoparticles), angular regions with rapidly varying $\delta$ (shown in **Figure 3b**) is accompanied with local maxima of $|\Delta_{GH}|$. Note that the spatial extent of the Poynting vector circulation ($P_\phi$ in Eq. 4) in scattering of LCP/RCP light may be extended to several λ for larger sized dielectric scatterer, leading to large SH shift [13]. In conformity with this, our calculations (for the dielectric microsphere with $a$ = 1μm, $n_s$ = 1.6) also yielded large magnitude of $\Delta_{SH}$ ($\geq \lambda$) at several narrow ranges of θ (not shown here). Importantly, the correlation between $|\Delta_{SH}(\theta)|$ and $|d(\theta)|$ was observed to remain intact. These results highlight the utility of the experimentally measurable retardance and *diattenuation* parameter of scattering for studying / interpreting GH and SH shift in scattering from micro/nano systems.

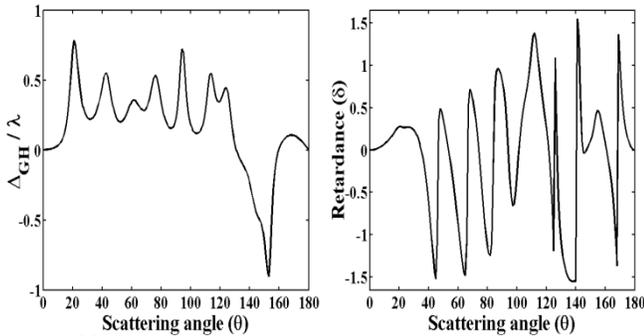

**Fig. 3. (a)** The computed GH shift $\Delta_{GH}$ as a function of θ for a dielectric microsphere (a = 1μm, refractive index = 1.6, λ = 632.8 nm). **(b)** The angular variation of *retardance $\delta$* (in radian) for the same dielectric scatterer.

To summarize, we have shown that the in-plane (longitudinal) Goos-Hänchen shift can also be observed in scattering of plane waves from micro/nano particles. Both the Goos-Hänchen and the Spin Hall shifts in plasmonic metal nanoparticles and dielectric micro-particles were studied utilizing the transverse components of the Poynting vector of the scattered wave. Among many interesting findings of such investigation, the notable ones are the following. While the magnitude of the GH shift from the dielectric particles are small (in the sub-wavelength domain), it exhibits giant enhancement in plasmonic metal nanoparticles. This giant enhancement of the GH shift originates from the interference of neighboring plasmon resonance modes. On the other hand, the influence of the interference of the plasmon modes on the SH shift is weaker, leading to relatively small magnitude of SH shift in plasmonic nanoparticles. Regarding experimental realization of the shifts, the measurement of the SH shift in scattering from dielectric microspheres has been realized by mapping the scattered intensity distributions for incident LCP/RCP states, in a direction transverse to the scattering plane [14]. The GH shift in scattering may also be measured in analogous fashion by mapping the differential scattered intensity distributions for incident orthogonal linear polarization states, in a direction parallel to a chosen scattering plane. Moreover, the observed correlations between the shifts and the *retardance* and the *diattenuation* parameters may open up interesting avenues for analysis / interpretation of the shifts via these experimentally measurable scattering polarimetry parameters.